\documentclass[showpacs,preprint,aps]{revtex4}%
\usepackage{amsfonts}
\usepackage{amsmath}
\usepackage{amssymb}
\usepackage{epsfig}
\usepackage{graphicx}%
\ifx\pdfoutput\relax\let\pdfoutput=\undefined\fi
\newcount\msipdfoutput
\ifx\pdfoutput\undefined\else
\ifcase\pdfoutput\else
\ifx\paperwidth\undefined\else
\ifdim\paperheight=0pt\relax\else\pdfpageheight\paperheight\fi
\ifdim\paperwidth=0pt\relax\else\pdfpagewidth\paperwidth\fi
\fi\fi\fi
\begin{document}
\title{Quantum control gates with weak cross-Kerr nonlinearity}
\author{Qing Lin}
\email{qlin@mail.ustc.edu.cn}
\affiliation{College of Information Science and Engineering, Huaqiao University (Xiamen),
Xiamen 361021, P.R.China.}
\author{Jian Li}
\affiliation{Department of Physics, Southeast University, Nanjing 211189, P. R. China}


\begin{abstract}
In this paper, with the weak cross-Kerr nonlinearity, we first present a
special experimental scheme called controlled-path gate with which the
realization of all possible bipartite positive-operator-value measurements of
two-photon polarization states may be nearly deterministic. Following the same
technique, the realization of quantum control gates, including the
controlled-NOT gate, Fredkin gate, Toffoli gate, arbitrary controlled-U gate,
and even arbitrary multi-controlled-U gate, are proposed. The corresponding
probabilities are 1/2, 1/8, 2/23, etc. respectively. Only the coherent states
are required but not any ancilla photons, and no coincidence measurement are
required which results in these gates are scalable. The structures of these
gates are very simple, and then we think they are feasible with the current
experimental technology in optics.

\end{abstract}
\maketitle

\section{Introduction}

In the quantum computation, quantum control gates play a very important role.
It was proven that two-qubit unitary gates and single-qubit gates are
sufficient for universal quantum computation \cite{Nielsen}. In linear optics,
many schemes are provided for the realization of two-qubit unitary gates, for
example, controlled-NOT (CNOT) gates \cite{CNOT} or controlled-phase gates
\cite{CP}. However, some of these gates work on the coincidence basis which
results in these gates are not scalable, \textit{i.e.}, these gates can not be
used to realize multi-qubit gates and then the universal computation.
Moreover, all these gates are probabilistic which result in the probability of
the realization of universal computation may be tiny, for the reason that many
two-qubit unitary gates required. For example, quantum Fredkin gate can be
constructed by five CNOT gates and some single-qubit gates \cite{Smolin}, and
the probability of CNOT gate is only $1/4$ in linear optics \cite{CNOT}, then
the probability of Fredkin gate is $4^{-5}=9.8\times10^{-4}$. To avoid the
inefficient, more efficient even deterministic gates must be looked for.
Fortunately, with the weak cross-Kerr nonlinearity, a parity projector
\cite{Barrett} and a deterministic CNOT gate \cite{Nemoto} had been proposed,
and then the universal computation can be realized deterministic in principle.
However, the universal computation and even a multi-qubit gate may be need too
many CNOT gates, then the structure may be too complex to be realized in
optics. Alternatively, it is interesting to look for some multi-qubit gates
with simple structure even though the probability is not unit. In this paper,
we will present the quantum control gates with very simple structure, and we
think these gates may be more feasible with the current experimental technology.

This paper is organized as follows. In sec.II, we first propose a scheme of a
gate we call it controlled-path (C-path) gate with the weak cross-Kerr
nonlinearities, and then use this gate to realize all possible bipartite
positive-operator-value measurements (POVMs) of two-photon polarization
states. In addition, this technique is developed to realize the CNOT gate,
Fredkin gate, Toffoli gate, controlled-U (CU) gate and even multi-controlled-U
(MCU) gate. Sec.III is for conclusion remarks.

\section{Quantum control gate}

Before we outline our schemes of quantum control gates, we briefly review the
useful weak cross-Kerr nonlinearity which has been used in Refs.
\cite{Barrett,Nemoto,Spiller,Kok}. Suppose a non-linear weak cross-Kerr
interaction between a signal state (photonic qubit) $\left\vert \psi
\right\rangle =c_{1}\left\vert 0\right\rangle +c_{2}\left\vert 1\right\rangle
+c_{2}\left\vert 2\right\rangle $ and a coherent state $\left\vert
\alpha\right\rangle $. After the evolution, the output state is,%

\begin{equation}
\left\vert \psi\right\rangle \left\vert \alpha\right\rangle \rightarrow
c_{1}\left\vert 0\right\rangle \left\vert \alpha\right\rangle +c_{2}\left\vert
1\right\rangle \left\vert \alpha e^{i\theta}\right\rangle +c_{2}\left\vert
2\right\rangle \left\vert \alpha e^{i2\theta}\right\rangle ,
\end{equation}
where $\theta$ is induced by the nonlinearity. Through a general
homodyne-heterodyne measurement of the phase of the coherent state, the signal
state $\left\vert \psi\right\rangle $ will be projected into a definite number
state or superposition of number states. Because the measurement can be
performed with high fidelity, the projection is nearly deterministic. This
technique has first been used to realize a parity projector \cite{Barrett},
and then a CNOT gate\cite{Nemoto}. It provides a new route to new quantum
computation \cite{Spiller}. The requirement for this technique is
$\alpha\theta>1$ \cite{Spiller}, where $\alpha$ is the amplitude of the
coherent state. Even with the weak nonlinearity ($\theta$ is small), this
requirement can be satisfied with large amplitude of the coherent state, then
this requirement may be feasible with current experimental technology. Our
schemes of quantum control gates also work with the weak cross-Kerr nonlinearity.

\subsection{C-path gate}

Firstly, we discuss the C-path gate. Here, we use the polarization of photons
as qubit and define the horizontally (vertically) linear polarization
$\left\vert H\right\rangle $($\left\vert V\right\rangle $) as the qubit
$\left\vert 0\right\rangle $($\left\vert 1\right\rangle $). Consider a
two-qubit initially prepared in the state $\left\vert \Psi\right\rangle
=\alpha\left\vert H\right\rangle _{1}\left\vert H\right\rangle _{2}%
+\beta\left\vert H\right\rangle _{1}\left\vert V\right\rangle _{2}%
+\gamma\left\vert V\right\rangle _{1}\left\vert H\right\rangle _{2}%
+\delta\left\vert V\right\rangle _{1}\left\vert V\right\rangle _{2}$, where
$\left\vert \alpha\right\vert ^{2}+\left\vert \beta\right\vert ^{2}+\left\vert
\gamma\right\vert ^{2}+\left\vert \delta\right\vert ^{2}=1$. In a C-path gate,
the paths of the first photon are controlled by the second photon. The
experimental setup is shown in Fig.1. The control photon is transmitted
through a balanced Mach-Zehnder (M-Z) interferometer formed by two polarizing
beam splitters (PBS$_{1}$, PBS$_{2}$) which let the photon $\left\vert
H\right\rangle $ be passed and the photon $\left\vert V\right\rangle $ be
reflected, while the target photon is injected into a 50:50 beam splitter
(BS). The two photons combined with a coherent state $\left\vert
\alpha\right\rangle $ interact with the cross-Kerr nonlinearities, such that a
phase shift will be induced in the coherent state. Suppose the control photon
induces a controlled phase shift $-\theta$, while the target photon induces a
controlled phase shift $\theta$, then the input state $\left\vert
\Psi\right\rangle \left\vert \alpha\right\rangle $ will evolve to the follows:%

\begin{align}
&  \frac{1}{\sqrt{2}}\left(  \alpha\left\vert H\right\rangle _{1}^{S_{1}%
}\left\vert H\right\rangle _{2}+\beta\left\vert H\right\rangle _{1}^{S_{2}%
}\left\vert V\right\rangle _{2}+\gamma\left\vert V\right\rangle _{1}^{S_{1}%
}\left\vert H\right\rangle _{2}+\delta\left\vert V\right\rangle _{1}^{S_{2}%
}\left\vert V\right\rangle _{2}\right)  \left\vert \alpha\right\rangle
\nonumber\\
&  +\frac{1}{\sqrt{2}}\left(  \alpha\left\vert H\right\rangle _{1}^{S_{2}%
}+\gamma\left\vert V\right\rangle _{1}^{S_{2}}\right)  \left\vert
H\right\rangle _{2}\left\vert \alpha e^{-i\theta}\right\rangle +\frac{1}%
{\sqrt{2}}\left(  \beta\left\vert H\right\rangle _{1}^{S_{1}}+\delta\left\vert
V\right\rangle _{1}^{S_{1}}\right)  \left\vert V\right\rangle _{2}\left\vert
\alpha e^{i\theta}\right\rangle ,
\end{align}
where the superscripts $S_{1}$, $S_{2}$ denote the paths of the first photon.
Through a general homodyne-heterodyne measurement (X homodyne measurement),
the two-photon state will be projected into the following state,
\begin{equation}
\alpha\left\vert H\right\rangle _{1}^{S_{1}}\left\vert H\right\rangle
_{2}+\beta\left\vert H\right\rangle _{1}^{S_{2}}\left\vert V\right\rangle
_{2}+\gamma\left\vert V\right\rangle _{1}^{S_{1}}\left\vert H\right\rangle
_{2}+\delta\left\vert V\right\rangle _{1}^{S_{2}}\left\vert V\right\rangle
_{2}.
\end{equation}
Here we only retain the case that no phase shift induced in the coherent
state, and the success probability is $P_{succ}^{CP}=1/2$. If a switch (S)
which will exchange the two photons and a phase shift conditionally controlled
by the homodyne detection through a classical feedforward are applied, this
C-path gate is nearly deterministic, i.e., $P_{succ,\max}^{CP}=1$. By the same
way, one can implement a multi-controlled-path gate in which multiple qubits
control the paths of the other qubits.

This C-path gate is very useful in the quantum computation for the reason that
many quantum control gates (for example, CNOT gate, Fredkin gate, etc.) can be
realized by some operations performed in the different paths of the target
photons. These schemes of quantum control gates will be discussed in the
following. Now we discuss the first use of this control-path gate. If we place
a half wave plate (HWP, set at 22.5$^{\circ}$-Hadamard gate) in the path of
the control photon which is shown in the dashed line of the Fig.1, the
following state can be achieved,%

\begin{equation}
\frac{1}{\sqrt{2}}\left(  \alpha\left\vert H\right\rangle _{1}^{S_{1}}%
+\beta\left\vert H\right\rangle _{1}^{S_{2}}+\gamma\left\vert V\right\rangle
_{1}^{S_{1}}+\delta\left\vert V\right\rangle _{1}^{S_{2}}\right)  \left\vert
H\right\rangle _{2}+\frac{1}{\sqrt{2}}\left(  \alpha\left\vert H\right\rangle
_{1}^{S_{1}}-\beta\left\vert H\right\rangle _{1}^{S_{2}}+\gamma\left\vert
V\right\rangle _{1}^{S_{1}}-\delta\left\vert V\right\rangle _{1}^{S_{2}%
}\right)  \left\vert V\right\rangle _{2}.
\end{equation}
If the detection of the control photon infers its polarization is $\left\vert
H\right\rangle $, the initial state $\left\vert \Psi\right\rangle $ has been
transferred onto the following state of a single photon in the Hilbert space
of its polarization and path states,%

\begin{equation}
\left\vert \Phi\right\rangle =\alpha\left\vert HS_{1}\right\rangle
+\beta\left\vert HS_{2}\right\rangle +\gamma\left\vert VS_{1}\right\rangle
+\delta\left\vert VS_{2}\right\rangle .
\end{equation}
The success probability is $P_{succ}^{CT}=1/2$. If a classical feedforward
phase shift $\pi$ is induced to the path $S_{2}$ when the detection infers the
polarization of the control photon is $\left\vert V\right\rangle $, the
success probability will increase to $1$.

The transformation $\left\vert \Psi\right\rangle \rightarrow\left\vert
\Phi\right\rangle $ is crucial for the realization of all possible bipartite
POVMs of two-photon polarization states in Ref. \cite{Ahnert}. In their
scheme, a special three-photon entangled state created by a quantum Fredkin
gate and a teleportation process of five photons are required for this
transformation. It is evident that our scheme is better than their scheme in
the amount of resource, the complexity of the operations, and the great
advantage of our scheme is the success probability is nearly unity which makes
the realization of all possible bipartite POVMs of two-photon polarization
states nearly deterministic.

\subsection{CNOT gate}

Secondly, we discuss the CNOT gate. Suppose two photons initially prepared in
the state $\left\vert \Psi\right\rangle $, and the CNOT gate can be described
by the following transformation,%

\begin{equation}
\left\vert \Psi\right\rangle \rightarrow\alpha\left\vert H\right\rangle
_{1}\left\vert H\right\rangle _{2}+\beta\left\vert H\right\rangle
_{1}\left\vert V\right\rangle _{2}+\gamma\left\vert V\right\rangle
_{1}\left\vert V\right\rangle _{2}+\delta\left\vert V\right\rangle
_{1}\left\vert H\right\rangle _{2},
\end{equation}
The experimental setup is shown in Fig.2, here the first photon is the control
photon which is transmitted through a balanced M-Z interferometer formed by
two PBSs (PBS$_{1}$, PBS$_{2}$), while the target photon is also transmitted
through a balanced\ M-Z interferometer formed by two BSs (BS$_{1}$, BS$_{2}$)
whose transmissivity (reflectivity)\ are $T_{1},T_{2}$ ($R_{1},R_{2}$)
respectively. A single-photon operation $\sigma_{x}$ is performed in one arm.
With the cross-Kerr nonlinearities and a X homodyne measurement associated
with the classical feedforward, the following states can be achieved in the output,%

\begin{equation}
\sqrt{T_{1}R_{2}}\left(  \alpha\left\vert H\right\rangle _{1}\left\vert
H\right\rangle _{2}+\beta\left\vert H\right\rangle _{1}\left\vert
V\right\rangle _{2}\right)  +\sqrt{R_{1}T_{2}}\left(  \gamma\left\vert
V\right\rangle _{1}\left\vert V\right\rangle _{2}+\delta\left\vert
V\right\rangle _{1}\left\vert H\right\rangle _{2}\right)  ,
\end{equation}
or%
\begin{equation}
\sqrt{R_{1}T_{2}}\left(  \alpha\left\vert H\right\rangle _{1}\left\vert
H\right\rangle _{2}+\beta\left\vert H\right\rangle _{1}\left\vert
V\right\rangle _{2}\right)  +\sqrt{T_{1}R_{2}}\left(  \gamma\left\vert
V\right\rangle _{1}\left\vert V\right\rangle _{2}+\delta\left\vert
V\right\rangle _{1}\left\vert H\right\rangle _{2}\right)  .
\end{equation}
Compared with the Eq. (6), it is immediately to find that the CNOT operation
is completed when the condition $\sqrt{T_{1}R_{2}}=\sqrt{R_{1}T_{2}}$ is
satisfied, and the success probability $P_{succ}^{CNOT}=2T_{1}R_{2}$. It is
easy to find that the maximum success probability is $P_{succ,\max}%
^{CNOT}=1/2$ when $T_{1}=R_{2}=1/2$. Compared with the scheme proposed by
Nemoto \textit{et al }\cite{Nemoto}, our scheme is probabilistic but no
ancilla photons are required.

\subsection{Fredkin gate}

Thirdly, we discuss the Fredkin gate which is also called controlled-swap
gate. Consider a single photon (control photon) in the state $\left\vert
\psi\right\rangle =\alpha\left\vert H\right\rangle +\beta\left\vert
V\right\rangle $ ($\left\vert \alpha\right\vert ^{2}+\left\vert \beta
\right\vert ^{2}=1$), and two photons (target photons) in the state
$\left\vert \phi\right\rangle =p_{1}\left\vert \Psi^{+}\right\rangle
+p_{2}\left\vert \Psi^{-}\right\rangle +p_{3}\left\vert \Phi^{+}\right\rangle
+p_{4}\left\vert \Phi^{-}\right\rangle $ ($\underset{i}{\sum}\left\vert
p_{i}\right\vert ^{2}=1$), where $\left\{  \left\vert \Psi^{\pm}\right\rangle
,\left\vert \Phi^{\pm}\right\rangle \right\}  $ are the Bell states. A Fredkin
gate can be described by the following transformation,%

\begin{align}
\left\vert \psi\right\rangle \left\vert \phi\right\rangle  &  \rightarrow
\alpha\left\vert H\right\rangle \left(  p_{1}\left\vert \Psi^{+}\right\rangle
+p_{2}\left\vert \Psi^{-}\right\rangle +p_{3}\left\vert \Phi^{+}\right\rangle
+p_{4}\left\vert \Phi^{-}\right\rangle \right) \nonumber\\
&  +\beta\left\vert V\right\rangle \left(  p_{1}\left\vert \Psi^{+}%
\right\rangle -p_{2}\left\vert \Psi^{-}\right\rangle +p_{3}\left\vert \Phi
^{+}\right\rangle +p_{4}\left\vert \Phi^{-}\right\rangle \right)  ,
\end{align}
that is, if the control photon is in the state $\left\vert H\right\rangle $,
the target two photons are unchanged; while the control photon is in the state
$\left\vert V\right\rangle $, a swap operation is implemented to the target
two photons. For the reason that, only the singlet state $\left\vert \Psi
^{-}\right\rangle $ is antisymmetric while the other three Bell states are
symmetric, the swap operation only results in a phase shift $\pi$ to the state
$\left\vert \Psi^{-}\right\rangle $ while the other states unchanged. Our
scheme of the Fredkin gate is shown in Fig.3. The control photon is
transmitted through a balanced M-Z interferometer formed by two PBSs
(PBS$_{1}$, PBS$_{2}$), while the two target photons are transmitted through a
balanced M-Z interferometer formed by two BSs (BS$_{1}$, BS$_{2}$ or BS$_{3}$,
BS$_{4}$) whose transmissivity (reflectivity)\ are $T_{1},T_{2}$ or
$T_{3},T_{4}$ ($R_{1},R_{2}$ or $R_{3},R_{4}$) respectively. In addition, a
balanced M-Z interferometer (in the dashed line of Fig.3) formed by two BSs
(BS$_{5}$, BS$_{6}$) associated with a phase shift $\pi$ in one arm is
required. The Hong-Ou-Mandel interference in this M-Z interferometer yields
the following transformation \cite{Hong}:%

\begin{equation}
\left\vert \Psi^{-}\right\rangle \rightarrow-\left\vert \Psi^{-}\right\rangle
;\text{ }\left\vert \Psi^{+}\right\rangle \left(  \left\vert \Phi^{\pm
}\right\rangle \right)  \rightarrow\left\vert \Psi^{+}\right\rangle \left(
\left\vert \Phi^{\pm}\right\rangle \right)  .
\end{equation}
Compared with the above two schemes, we change the phase shift induced by the
control photon to be $-2\theta$, while the phase shift is $\theta$ for the two
target photons. If the cross-Kerr nonlinearities are used and we retain the
case that no phase shift induced in the coherent state, we will achieve the
following state in the output:%

\begin{align}
&  \sqrt{T_{1}R_{2}R_{3}T_{4}}\alpha\left\vert H\right\rangle \left(
p_{1}\left\vert \Psi^{+}\right\rangle +p_{2}\left\vert \Psi^{-}\right\rangle
+p_{3}\left\vert \Phi^{+}\right\rangle +p_{4}\left\vert \Phi^{-}\right\rangle
\right)  \nonumber\\
&  +\sqrt{R_{1}T_{2}T_{3}R_{4}}\beta\left\vert V\right\rangle \left(
p_{1}\left\vert \Psi^{+}\right\rangle -p_{2}\left\vert \Psi^{-}\right\rangle
+p_{3}\left\vert \Phi^{+}\right\rangle +p_{4}\left\vert \Phi^{-}\right\rangle
\right)  .
\end{align}
Compared with the Eq.(9), the Fredkin gate is realized when the condition
$\sqrt{T_{1}R_{2}R_{3}T_{4}}=\sqrt{R_{1}T_{2}T_{3}R_{4}}$ is satisfied, then
the success probability is $P_{succ}^{Fredkin}=T_{1}R_{2}R_{3}T_{4}$. Hence
the maximum success probability is $P_{succ}^{Fredkin}=1/16$ when $T_{1}%
=R_{2}=R_{3}=T_{4}=1/2$. Moreover, if a M-Z interferometer which is same to
the M-Z interferometer in the dashed line associated with a phase shift $\pi$
conditionally controlled by the homodyne detection (the phase of the coherent
state is $\pm2\theta$) through a classical feedforward, is implemented in the
outputs of BS$_{2}$ and BS$_{4}$, the probability may be $P_{succ,\max
}^{Fredkin}=1/8$.

Now we compare our scheme of Fredkin gate with the previous schemes. In 1989,
Milburn used the cross-Kerr nonlinearities to realize the Fredkin gate
\cite{Milburn}, however, its cross-Kerr nonlinearities operate in single
photon level, so it requires huge nonlinearities which is a great challenge
for the current experimental technology. In linear optics, two types of
Fredkin gate, heralded gate and post-selected gate, had been proposed
\cite{Gong, Fiurasek1, Fiurasek2}. Exclusive of the requirement of ancilla
photons and small probability, the shortcomings of these gates are obvious.
The heralded Fredkin gates require single-photon detectors which is also a
great challenge for the current technology, and the post-selected Fredkin
gates work on the coincidence basis which results in these gates are not
scalable. Compared with these schemes, only the coherent states are required
in our scheme, and the structure is so simple that we think it is feasible
with the current technology.

\subsection{Toffoli gate, CU gate and MCU gate}

A little change that a CNOT gate or arbitrary two-qubit unitary gate replaces
the setups in the dashed line of the Fig.3, associated with appropriate
transmissivities of the four beam splitters, is enough for the realization of
the Toffoli gate or the CU gate. In the following, we calculate the
probability of Toffoli gate and the CU gate. For the Toffoli gate, two
coherent states are required because a CNOT gate is included in this scheme.
Consider a single photon (control photon) in the state $\left\vert
\psi\right\rangle =\alpha\left\vert H\right\rangle +\beta\left\vert
V\right\rangle $ ($\left\vert \alpha\right\vert ^{2}+\left\vert \beta
\right\vert ^{2}=1$), and two photons (target photons) in the state
$\left\vert \phi\right\rangle =q_{1}\left\vert HH\right\rangle +q_{2}%
\left\vert HV\right\rangle +q_{3}\left\vert VH\right\rangle +q_{4}\left\vert
VV\right\rangle $ ($\underset{i}{\sum}\left\vert q_{i}\right\vert ^{2}=1$).
Suppose that the transmissivities (reflectivities)\ of the four BSs are
$T_{1},T_{2},T_{3},T_{4}$ ($R_{1},R_{2},R_{3},R_{4}$) respectively, now the
modified scheme of the Fredkin gate will evolve the initial state $\left\vert
\psi\right\rangle \left\vert \phi\right\rangle $ to the follows (here we also
retain the case that no phase shift induced in the coherent state),%

\begin{align*}
&  \sqrt{T_{1}R_{2}R_{3}T_{4}}\alpha\left\vert H\right\rangle \left(
q_{1}\left\vert HH\right\rangle +q_{2}\left\vert HV\right\rangle
+q_{3}\left\vert VH\right\rangle +q_{4}\left\vert VV\right\rangle \right)  \\
&  +\frac{1}{\sqrt{2}}\sqrt{R_{1}T_{2}T_{3}R_{4}}\beta\left\vert
V\right\rangle \left(  q_{1}\left\vert HH\right\rangle +q_{2}\left\vert
HV\right\rangle +q_{3}\left\vert VV\right\rangle +q_{4}\left\vert
VH\right\rangle \right)  ,
\end{align*}
where the coefficient $1/\sqrt{2}$ is induced by the CNOT gate. The Toffoli
gate is completed when the condition $\sqrt{T_{1}R_{2}R_{3}T_{4}}=\frac
{1}{\sqrt{2}}\sqrt{R_{1}T_{2}T_{3}R_{4}}$ is satisfied. The success
probability is $P_{succ}^{Toffoli}=T_{1}R_{2}R_{3}T_{4}$. Choose $T_{1}%
=R_{2}=R_{3}=T_{4}=\frac{1}{\sqrt[4]{2}+1}$, the success probability may be
$P_{succ}^{Toffoli}\doteq\frac{1}{23}$. Similarly, a CNOT gate conditional
controlled by the homodyne detection (the phase of the coherent state is
$\pm2\theta$) through a classical feedforward is implemented in the outputs of
BS$_{2}$ and BS$_{4}$, the probability may be $P_{succ,\max}^{Toffoli}%
=\frac{2}{23}$.

In linear optics, two types of Toffoli gate, heralded gate and post-selected
gate, had been proposed \cite{Fiurasek2, Ralph}. Similarly, exclusive of the
requirement of ancilla photons and small probability, the uses of
single-photon detectors and the coincidence measurement limit their use in the
universal computation. These shortcomings are not exist in our scheme, and the
simple structure makes it much feasible with current technology.

The realization of CU gate is similar, and the success probability is
determined by the probability of the arbitrary unitary gate (suppose as $1/p$)
which can be realized by some CNOT gates and single-qubit gates, and the
transmissivities of the four beam splitters. The condition for the CU gate is
$\sqrt{T_{1}R_{2}R_{3}T_{4}}=\frac{1}{\sqrt{p}}\sqrt{R_{1}T_{2}T_{3}R_{4}}$.
Also choose $T_{1}=R_{2}=R_{3}=T_{4}=\frac{1}{\sqrt[4]{p}+1}$, the success
probability of CU gate may be $P_{succ}^{CU}=\left(  \frac{1}{\sqrt[4]{p}%
+1}\right)  ^{4}$, and it may be $P_{succ,\max}^{CU}=2\left(  \frac
{1}{\sqrt[4]{p}+1}\right)  ^{4}$ with some additional setups similar to the
Toffoli gate. In addition, it is straightforward to develop this technique to
the realization of MCU gate which is shown in Fig.4. The realization is
described in the following. The control photons are all transmitted through a
balanced M-Z interferometer formed by two PBSs respectively, while the target
photons are all transmitted through a balanced M-Z interferometer formed by
two BSs respectively. Next, similar to the setups in the dashed line of Fig.3,
in one arm of all the M-Z interferometers form by the BSs, we implement a
multi-qubit unitary gate which can be realized by the quantum control gates
described above. Assisted by some coherent states and the weak cross-Kerr
nonlinearity, the MCU gate can be realized associated with the appropriate
transmissivities of the BSs. Compared with the realization of MCU gate with
many CNOT gates and single-qubit gates, our scheme can reduce the complexity
of the realization greatly.

\section{Conclusion}

In this paper, with the weak cross-Kerr nonlinearity, we first present a
special experimental scheme called C-path gate with which the realization of
all possible bipartite POVMs of two-photon polarization states can be simpler
and nearly deterministic. Following the same technique, the schemes of the
realization of quantum control gates have been proposed, including the CNOT
gate $\left(  1/2\right)  $, Fredkin gate $\left(  1/8\right)  $, Toffoli gate
$\left(  2/23\right)  $, CU gate and even MCU gate. All these gates are
scalable with the certain probabilities which are larger than those gates in
linear optics. Less resource are required and the structures of these gates
are so simple that we think they are feasible with current technology and may
be useful for the realization of universal computation in optics.

\begin{acknowledgments}
The author would like to thank Dr Pieter Kok and Bill Munro for
their helpful discussions. The author Qing Lin was funded by the
HuaQiao University Foundation, China (Grant No. 07BS406).
\end{acknowledgments}

\begin{figure}[tbh]
\begin{center}
\epsfig{file=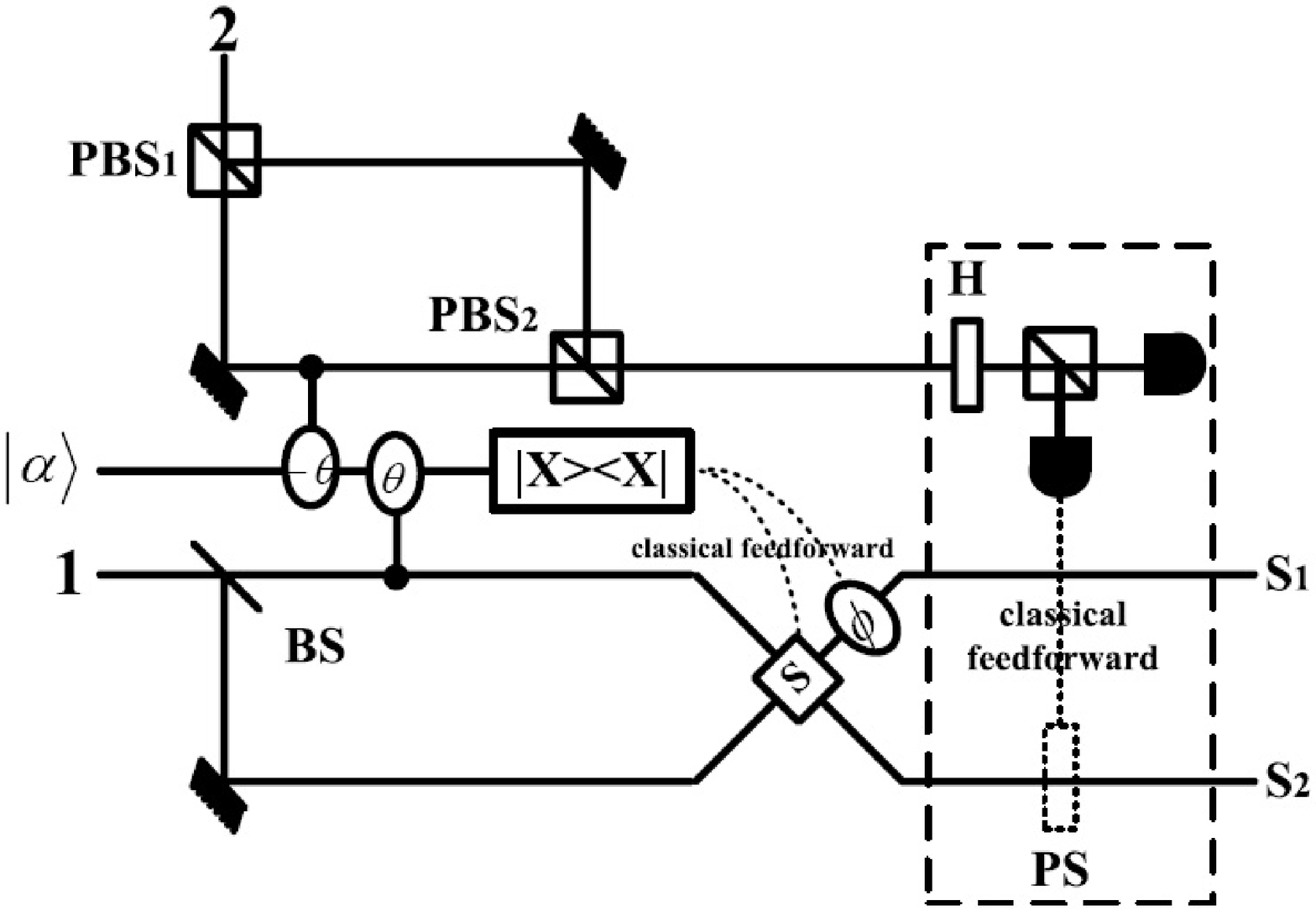,width=13cm}
\end{center}
\caption{Controlled-path gate with weak cross-Kerr nonlinearity. Assisted by
the switch (S) and the phase shift conditional controlled by the homodyne
detection through a classical feedforward, this gate is nearly deterministic.
If the setups in the dashed line are used, this gate can be used to realize
all possible bipartite positive-operator-value measurements of two-photon
polarization states nearly determinately.}%
\end{figure}

\begin{figure}[tbh]
\begin{center}
\epsfig{file=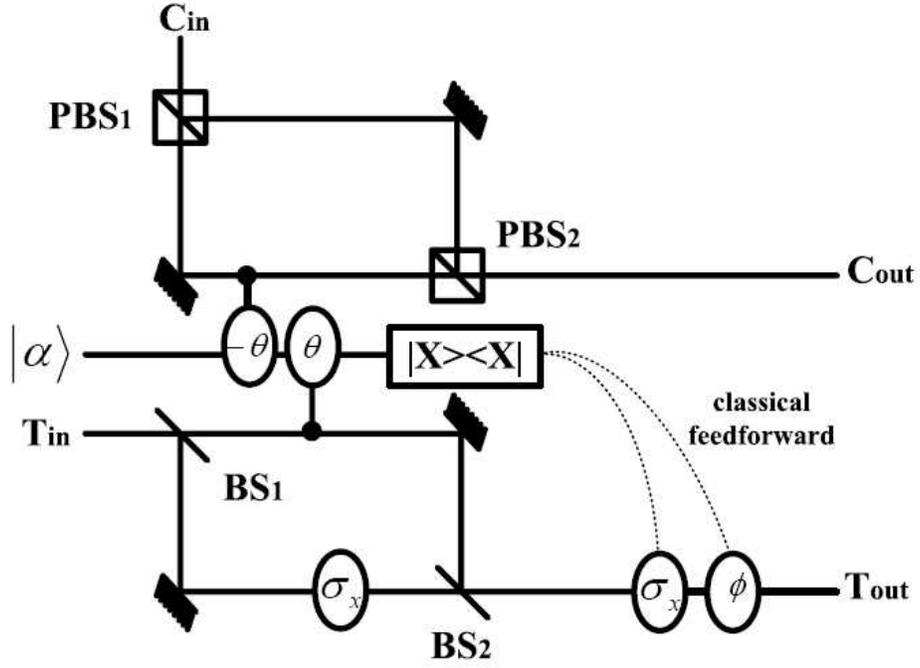,width=13cm}
\end{center}
\caption{CNOT gate with the weak corss-Kerr nonlinearity. Assisted by a
classical feedforward, this gate can be implemented with the probability 1/2.}%
\end{figure}

\begin{figure}[tbh]
\begin{center}
\epsfig{file=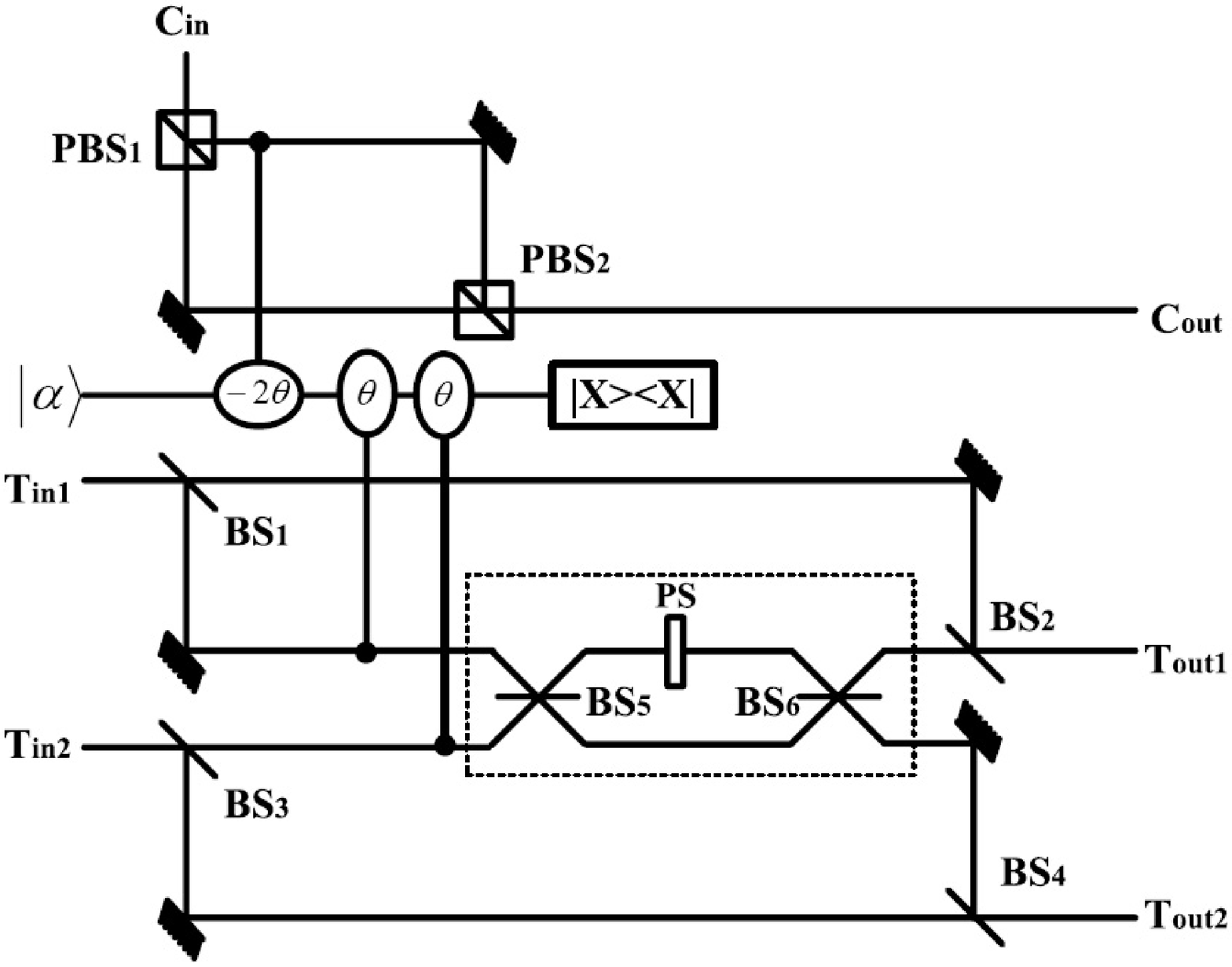,width=13cm}
\end{center}
\caption{Fredkin gate with the weak cross-Kerr nonlinearities. The setups in
the dashed line will complete the transformation $\left\vert \Psi
^{-}\right\rangle \rightarrow-\left\vert \Psi^{-}\right\rangle ;\left\vert
\Psi^{+}\right\rangle \left(  \left\vert \Phi^{\pm}\right\rangle \right)
\rightarrow\left\vert \Psi^{+}\right\rangle \left(  \left\vert \Phi^{\pm
}\right\rangle \right)  .$ Associated with the nonlinearities and appropriate
transmissivities of four beam splitters, the Fredkin gate is realized with the
probability 1/16. If some additional setups are used, the probability will
increase to 1/8. For details, see text.}%
\end{figure}

\begin{figure}[tbh]
\begin{center}
\epsfig{file=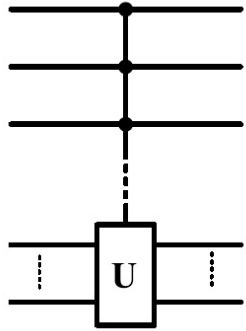,width=4cm}
\end{center}
\caption{Multi-control-U gate. }%
\end{figure}

\end{document}